%% file: ms.tex
\documentclass[preprint]{acmart}

\usepackage{amsmath}
\usepackage{hyperref}
\usepackage{graphicx}
\usepackage{todonotes}
\usepackage{xspace}
\usepackage[utf8]{inputenc}
\usepackage{mdframed}
\usepackage{ulem}
\usepackage{enumitem}
\usepackage{amsthm}
\usepackage[many]{tcolorbox}
\usepackage{multirow}

\newcommand{\myparagraph}[1]{\paragraph{\uline{#1}}}

\input{commands}
\input{relalg-packages}

\input{relalg-commands}
\input{listings}

\acmDOI{10.475/123_4}

\acmISBN{123-4567-24-567/08/06}

\acmConference[SIGMOD SRC]{SIGMOD ACM Student Research Competition}{June 2018}{Houston, TX} 
\acmYear{2018}
\copyrightyear{2018}

\acmPrice{15.00}

\begin{document}

\title{Incremental View Maintenance for Property Graph Queries}

\author{G\'abor Sz\'arnyas}
\affiliation{%
	\institution{Budapest University of Technology and 
	Economics\\Department of Measurement and Information Systems\\MTA-BME 
	Lend\"ulet Research Group on Cyber-Physical Systems}
}
\email{szarnyas@mit.bme.hu}

\maketitle

\input{problem-and-motivation}

\input{preliminaries}

\input{related-work}

\input{approach-and-contributions}

\clearpage

\bibliographystyle{ACM-Reference-Format}
\bibliography{bib,szarnyasg}

\end{document}

%% file: commands.tex
\presetkeys{todonotes}{inline}{}

\newcommand{\viatra}{\mbox{\textsc{Viatra}}\xspace}

\newcommand{\ie}{i.e.\@\xspace}

\newcommand{\eg}{e.g.\@\xspace}

\newcommand{\opencypher}{openCypher\xspace}

\newcommand{\oci}{\mbox{OCI}\xspace}

\theoremstyle{rqtheorem}
\newtheorem*{rqtheorem*}{Research Question}

\tcolorboxenvironment{rqtheorem*}{
	sharp corners,
	boxrule=0.6pt,
	colback=white,
	before skip=2pt,
	after skip=2pt,
	boxsep=-2pt,
}

\theoremstyle{rstheorem}
\newtheorem*{rstheorem*}{Research Statement}

\tcolorboxenvironment{rstheorem*}{
	sharp corners,
	boxrule=0.6pt,
	colback=white,
	before skip=2pt,
	after skip=2pt,
	boxsep=-2pt,
}

\newcommand{\vertexlabels}{L}
\newcommand{\edgelabels}{T}

\newcommand{\verticestoedges}{\mathit{st}}

\newcommand{\vertexproperties}{P_v}
\newcommand{\edgeproperties}{P_e}

\newcommand{\vertexlabelfunction}{\mathcal{L}}
\newcommand{\edgelabelfunction}{\mathcal{T}}

\newcommand{\dom}[1]{\mathrm{dom}(#1)}

%% file: relalg-packages.tex
\usepackage{ifxetex}
\usepackage{ifluatex}
\newif\ifxetexorluatex 
\ifnum 0\ifxetex 1\fi\ifluatex 1\fi>0
   \xetexorluatextrue
\fi

\ifxetexorluatex
  \usepackage{fontspec}
\else
  \usepackage[T1]{fontenc}
  \usepackage[utf8]{inputenc}
\fi

\usepackage{amsmath}
\usepackage{amssymb}
\usepackage{etoolbox}
\usepackage{xspace}
\newtoggle{textualoperators}

\usepackage{tikz}
\usepackage{forest}

\tikzset{every node/.style={draw}}

%% file: relalg-commands.tex
\newcommand{\assign}{\rightarrow}

\newcommand{\atom}[1]{\mathsf{#1}}

\newcommand{\colonseparator}{:}

\newcommand{\var}[1]{\mathtt{#1}}
\newcommand{\edgevariable}[2]{\var{#1}\ifstrempty{#2}{}{\hspace{-0.6ex}\colonseparator\hspace{-0.6ex}{\atom{#2}}}}
\newcommand{\vertexvariable}[2]{(\var{#1}\ifstrempty{#2}{}{\colonseparator{\atom{#2}}})}

\newcommand{\relalgop}[1]{\textsc{#1}}

\newcommand{\getverticesop}{\iftoggle{textualoperators}{\relalgop{GetVertices}
	}{
		\bigcirc
	}}

\newcommand{\getedgesopdirected}{\iftoggle{textualoperators}{
		\relalgop{GetEdges}
	}{
		\Uparrow
	}}

\newcommand{\expandoutop}{\iftoggle{textualoperators}{
		\relalgop{ExpandOut}
	}{
		\uparrow
	}}

\newcommand{\projectionop}{\iftoggle{textualoperators}{
		\relalgop{Projection}
	}{
		\pi
	}}

\newcommand{\selectionop}{\iftoggle{textualoperators}{
		\relalgop{Selection}
	}{
		\sigma
	}}

\newcommand{\unnestop}{\iftoggle{textualoperators}{
		\relalgop{Unnest}
	}{
		\mu
	}}

\newcommand{\joinop}{\iftoggle{textualoperators}{
		\relalgop{Join}
	}{
		\bowtie
	}}

\newcommand{\unionop}{\iftoggle{textualoperators}{
		\relalgop{Union}
	}{
		\cup
	}}

\newcommand{\cartesianproductop}{\iftoggle{textualoperators}{
		\relalgop{CartesianProduct}
	}{
		\times
	}}

\newcommand{\getvertices}[2]{\getverticesop_{\vertexvariable{#1}{#2}}}
\newcommand{\getedgesdirected}[6]{\getedgesopdirected_{\vertexvariable{#1}{#2}}^{\vertexvariable{#3}{#4}} \left[ \edgevariable{#5}{#6} \right]}

\newcommand{\kleenestar}{\ast}
\newcommand{\nagivationbody}[3]{\,_{\vertexvariable{#1}{}}^{\vertexvariable{#2}{#3}}}
\newcommand{\expandedgevariable}[4]{
	\left[
	\edgevariable{#1}{#2}
	\ifstrequal{#3}{1} 
	{
		\ifstrequal{#4}{1}
		{} 
		{\kleenestar_\atom{#3}^\atom{#4}} 
	} 
	{\kleenestar_\atom{#3}^\atom{#4}}
	\right]}

\newcommand{\expandout}[7]{\expandoutop \nagivationbody{#1}{#2}{#3} \expandedgevariable{#4}{#5}{#6}{#7} }

\newcommand{\projection}[2]{\projectionop_{#1}^{#2}}

\newcommand{\selection}[1]{\selectionop_{#1}}

\newcommand{\unnest}[1]{\unnestop_{#1}}

\newcommand{\join}{\joinop}

\newcommand{\union}{\unionop}

\newcommand{\cartesianproduct}{\cartesianproduct}

\definecolor{red}{HTML}{e41a1c}
\definecolor{blue}{HTML}{377eb8}
\definecolor{green}{HTML}{4daf4a}
\definecolor{lilac}{HTML}{984ea3}

\definecolor{progressbargreen}{HTML}{008000}

\newcommand{\externalschemacolorname}{red}
\newcommand{\extravariablescolorname}{blue}
\newcommand{\internalschemacolorname}{green}

\colorlet{externalschemacolor}{\externalschemacolorname}
\colorlet{extravariablescolor}{\extravariablescolorname}
\colorlet{internalschemacolor}{\internalschemacolorname}
\colorlet{nullarynodecolor}{lilac}

\newcommand{\operatortext}[1]{#1\xspace}

\newcommand{\getverticestext}{\operatortext{get-vertices}}
\newcommand{\getedgestext}{\operatortext{get-edges}}

\newcommand{\expandouttext}{\operatortext{expand-out}}

\newcommand{\schema}[1]{\mathrm{sch}(\mathit{#1})}

\newcommand{\grv}{\mathit{\alpha}}
\newcommand{\gre}{\mathit{\beta}}

%% file: listings.tex
\usepackage{listings}

\usepackage{textcomp}
\definecolor{keyword}{HTML}{2771a3}
\definecolor{pattern}{HTML}{b53c2f}
\definecolor{string}{HTML}{be681c}
\definecolor{relation}{HTML}{7e4894}
\definecolor{variable}{HTML}{107762}
\definecolor{comment}{HTML}{8d9094}

\lstdefinelanguage{cypher}
{
	morekeywords={
		MATCH, OPTIONAL, WHERE, NOT, AND, OR, XOR, RETURN, DISTINCT, ORDER, BY, ASC, ASCENDING, DESC, DESCENDING, UNWIND, AS, UNION, WITH, ALL, CREATE, DELETE, DETACH, REMOVE, SET, MERGE, SET, SKIP, LIMIT, IN,
		INDEX, DROP, UNIQUE, CONSTRAINT, EXPLAIN, PROFILE, START, CASE,
		GROUP, HAVING,
	},
	sensitive=true,
	morecomment=[l]{//},
	morecomment=[s]{/*}{*/},
	morestring=[b]{"},
}

\newcommand{\mycdots}{\cdot\!\cdot\!\cdot}

%% file: problem-and-motivation.tex
\vspace{-3ex}
\section{Problem and Motivation}
\label{sec:problem-and-motivation}

Graph processing challenges are common in modern database systems, with the property graph data model gaining widespread adoption~\cite{DBLP:journals/corr/abs-1709-03188}. Due to the novelty of the field, graph databases and frameworks typically provide their own query language, such as Cypher for Neo4j~\cite{GraphDatabases}, Gremlin for TinkerPop~\cite{DBLP:conf/dbpl/Rodriguez15} and GraphScript for SAP HANA~\cite{DBLP:conf/dbpl/ParadiesKBFKG17}. These languages often lack a formal background for their data model and semantics~\cite{DBLP:journals/corr/AnglesABHRV16}. To address this, the \opencypher initiative~\cite{openCypher} aims to standardise a subset of the Cypher language, for which it currently provides grammar specification and a set of acceptance tests to allow vendors to implement their \opencypher compatible engine.

Incremental view maintenance has been used for decades in relational database systems~\cite{DBLP:conf/sigmod/BlakeleyLT86}. In the graph domain, numerous use cases rely on complex queries and require low latency, including financial fraud detection, source code analysis~\cite{DBLP:journals/infsof/UjhelyiSHCVVF15} and checking integrity (or well-formedness) constraints in databases~\cite{TrainBenchmarkSOSYM}. While these could benefit from incremental evaluation, currently no property graph system provides incremental views.
Our research investigates the incremental view maintenance for \opencypher queries. 
A key challenge is that the property graph data model includes lists and maps, and queries can return arbitrarily nested data structures.

We propose three desirable properties for an incremental property graph query engine:
(IVM) incremental view maintenance, (FGN) fine granularity update operations on nested data structures, (ORD) ordering.  Previous research showed that IVM and FGN is possible~\cite{DBLP:conf/ideas/LiuVM99}. However, as stated in~\cite{XQueryTechReport}, \emph{"incremental view maintenance [IVM] strategies for data models that preserve order [ORD] remain an open problem to date"}. While removing support for \emph{ordering} might seem a plausible workaround, it would pose serious limitations: (1) queries that require top-$k$ results are common~\cite{LDBC_SNB} and (2) even more importantly, Cypher handles paths as an alternating list of vertices and edges, which must be kept ordered.
Therefore, we investigate the following research question:
\emph{Which practical fragment of the \opencypher language is incrementally} 
maintainable?

%% file: preliminaries.tex
\section{Preliminaries}
\label{sec:preliminaries}

\myparagraph{Data model}

A \emph{property graph} is $G = (V, E, \verticestoedges, \vertexlabels, \edgelabels, \vertexlabelfunction, \edgelabelfunction, \vertexproperties, \edgeproperties)$, where $V$ is a set of vertices, $E$ is a set of edges and $\verticestoedges: E \assign V \cartesianproductop V$ assigns the source and target vertices to edges. Vertices are labelled from $\vertexlabels$ by function $\vertexlabelfunction$ and edges are typed from $\edgelabels$ by function $\edgelabelfunction$. Let $D = \cup_{i} D_i$ be the union of atomic domains $D_i$. $\vertexproperties$ is a set of vertex properties. A vertex property $p_i \in \vertexproperties$ is a partial function $p_i: V \assign D_i$. 
Edge properties $P_e$ can be defined similarly.

Given a property graph $G$, relation $r$ is a \emph{graph relation} if the following holds~\cite{DBLP:conf/edbt/HolschG16}: $\forall A \in \schema{r}: \dom{A} \subseteq V \union E \union D,$
where $\schema{r}$ is the schema of $r$ (a list containing attribute names), $\dom{A}$ is the domain of attribute $A$, and $V$/$E$ are the vertices/edges of $G$.

\myparagraph{Running example}

We use the following example graph:

\begin{center}
\includegraphics[scale=0.45]{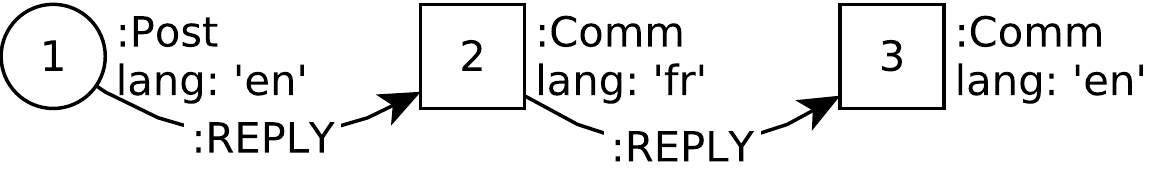}
\end{center}
\vspace{-1ex}

We use an example query that lists \textsf{Post}s $p$, along with threads $t$ that contain (transitive) reply \textsf{Comm[ent]}s that are written in the same lang[uage] as the \textsf{Post}. The result is shown on the right. (For conciseness, edges are omitted from paths throughout the paper.)

\vspace{1ex}
\begin{center}
\begin{minipage}{5.4cm}
\begin{lstlisting}
MATCH t = (p:Post)-[:REPLY*]->(c:Comm)
WHERE p.lang = c.lang
RETURN p, t
\end{lstlisting}
\end{minipage}
\quad
\begin{minipage}{2cm}
	\footnotesize
	\vspace{-4.8ex}
	\begin{tabular}{|l|l|}
		\hline
		\bf p & \bf t \\ \hline\hline
		1     & $[1, 2]$    \\ \hline
		1     & $[1, 2, 3]$ \\ \hline
	\end{tabular}
\end{minipage}
\end{center}

\myparagraph{GRA}

Graph queries can be formulated in \emph{graph relational algebra} (GRA)~\cite{DBLP:conf/adbis/MartonSV17}, which introduces two graph-specific operators: (1)~the \emph{\getverticestext} nullary operator $\getvertices{v}{V}$, which returns vertices $v$ with a label $V$ to serve as a base relation for later operators, (2)~the \emph{\expandouttext} unary operator $\expandout{v}{W}{W}{}{E}{1}{1} (r)$ that navigates from $v$ on an edge typed $E$ to a vertex $w$ with label $W$. The \expandouttext operator can also define transitive closure patterns, denoted by the $\ast$ symbol. GRA allows nested data structures, \ie if $x$ is an attribute of a graph relation, $x.p$ accesses the value of property $p$ in $x$~\cite{DBLP:conf/edbt/HolschG16}.

\myparagraph{NRA}

To allow precise formalisation of nested data structures, we use \emph{nested relational algebra} (NRA)~\cite{DBLP:conf/pods/JaeschkeS82,DBLP:journals/tcs/Bussche01}, which allows arbitrary nesting of relations. To access nested values, attribute $A$ of a nested relation $r$ can be \emph{unnested} using the operator $\unnest{A} (r)$. Nested relations can also represent properties of vertices/edges along with collections such as lists and maps.
We present two nested relations $\grv$ and $\gre$ that store the vertices and edges of the graph, respectively:

\begin{center}
\footnotesize
\begin{minipage}[b]{4cm}
	$\grv$
	\begin{tabular}{|l|l|l|}
		\hline
		\bf id & \bf label & \bf properties \\ \hline\hline
		1 & Post &
		\begin{tabular}{|l|l|}
			\hline
			\bf key & \bf value  \\ \hline
			lang    & en         \\ \hline
		\end{tabular} \\ \hline
		2 & Comm & \ldots \\ \hline
	\end{tabular}
\end{minipage}
\quad
\begin{minipage}[b]{4cm}
 	$\gre$
	\begin{tabular}{|l|l|l|l|}
		\hline
		\bf \bf s & \bf t & \bf type & \bf properties \\ \hline\hline
		1 & 2 & REPLY & \\ \hline
		2 & 3 & REPLY & \\ \hline
	\end{tabular}
\end{minipage}
\end{center}

\noindent We define operators formally as $\getvertices{v}{V} \equiv \projection{\atom{id} \assign \atom{v}}{} \selection{\grv.\atom{label} = \atom{V}} (\grv)$ and \\
$\expandout{v}{}{W}{}{E}{1}{1} (r) \equiv \selection{r.v = \gre.\atom{s} \land \gre.\atom{type} = \atom{E} \land \gre.\atom{t} = \grv.\atom{id} \land \grv.\atom{label} = \atom{W}} \left( r \join \gre \join \grv \right).$

\clearpage

%% file: related-work.tex
\section{Related Work}
\label{sec:related-work}

\myparagraph{Cypher}
Due to its novelty, there are only a few research works on the formalisation of (open)Cypher. An early attempt to provide a framework for the theoretical representation of \opencypher queries was published in~\cite{DBLP:conf/edbt/HolschG16}. In~\cite{DBLP:conf/adbis/MartonSV17}, we published a formalisation of a subset of \opencypher that mapped queries to GRA. The \emph{Cypher for Apache Spark} project is an ongoing effort to adapt the Cypher language to Spark~\cite{CAPS}. None of these works consider IVM.
Graphflow~\cite{DBLP:conf/sigmod/KankanamgeSMCS17} is an active graph database for incremental \opencypher queries. However, it does not support nested data structures.


\myparagraph{IVM of graph queries}
The \viatra framework~\cite{DBLP:journals/sosym/VarroBHHRU16} provides an incremental query engine over the object-oriented Eclipse Modeling Framework. However, it does not support FGN or ORD.
Strider~\cite{DBLP:journals/pvldb/RenCKLBRZK17} is a system supporting continuous SPARQL queries. As the RDF data model does not handle collections as first class citizens (only head-tail style lists are supported), FGN is not supported. 

\myparagraph{Querying nested data structures}
Paper~\cite{DBLP:journals/tkde/KunoR98} presents a method for incremental view maintenance in object-oriented databases, but ordering is not supported. Recently, the authors of~\cite{DBLP:conf/dlog/BotoevaCCRX16,DBLP:journals/corr/BotoevaCCRX16} formalised the language of the \mbox{MongoDB} document store using nested relational algebra, including ordering. However, IVM was not considered. An approach for incremental calculation of XQuery expressions is presented in~\cite{DBLP:conf/er/DimitrovaER03} and its accompanying technical report~\cite{XQueryTechReport}.

%% file: approach-and-contributions.tex
\section{Approach and Contributions}
\label{sec:approach-and-contributions}

As discussed in \autoref{sec:problem-and-motivation}, order-preserving lists are required to store paths. Henceforth, we propose a property graph query model that only allows (unordered) \emph{bags}, except for paths that are still stored \emph{as a list}, but can only be updated as an atomic unit (\ie the previous path has to be deleted and the new one has to be inserted).
We argue that the distinction between collection properties and paths makes sense from a practical point of view: collection properties often receive updates, while paths only benefit from incremental updates in rare cases (\eg when a single transaction deletes an edge in the path, but adds another one that keeps the path from deleting).

\myparagraph{Overview}

We propose the following workflow for compiling property graph queries to an incrementally maintainable expression and use the example of \autoref{sec:preliminaries} for illustration.

\begin{center}
\includegraphics[scale=0.45]{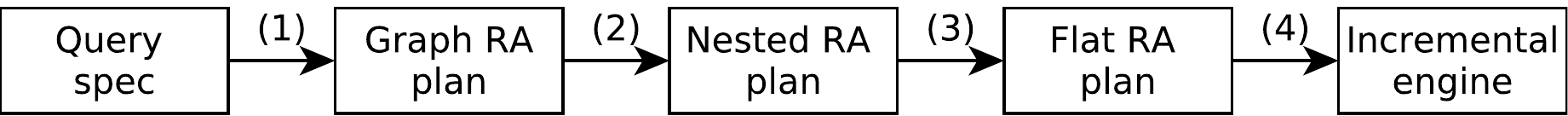}
\end{center}
\vspace{-0.7ex}

(1)~Compile the queries to GRA. A mapping from \opencypher was given in our earlier work~\cite{DBLP:conf/adbis/MartonSV17}. The example query results in:
$$  
\projection{\atom{p, t}}{}
\selection{\atom{c.lang} = \atom{p.lang}}
\left( 
\expandout{p}{c}{Comm}{}{REPLY}{}{}
\left(
\getvertices{p}{Post}
\right)
\right)
$$

(2)~Transform GRA to NRA, which is the key step to allow incremental maintenance. As expand operators cannot be maintained incrementally, they are replaced with joins. For this, we introduce the nullary \emph{\getedgestext} operator $\getedgesdirected{v}{V}{w}{W}{e}{E}$ that returns triples $(v, e, w)$ for each edge $e$ of type $E$ between $v$ of label $V$ and $w$ of label $W$.
Using this, each \emph{\expandouttext} is replaced with \emph{natural joins}:
$$\expandout{v}{w}{W}{}{E}{1}{1} (r) \equiv r \join \getedgesdirected{v}{V}{}{E}{w}{W}$$
Similarly, \emph{transitive \expandouttext}s are replaced with \emph{transitive joins}: $$\expandout{v}{w}{W}{}{E}{}{} (r) \equiv r \join^\ast \getedgesdirected{v}{V}{}{E}{w}{W}$$

Unlike relational databases, property graphs do not have a predefined schema. Hence, we slightly modify the \emph{unnest} operator (\autoref{sec:preliminaries}) so that defines specific attribute(s) to be unnested from the nested relation. For example, $\unnest{\atom{c.lang} \assign \atom{cL}}$ extracts the $\atom{lang}$ property of $\atom{c}$.
Using these rules, the example is transformed to:
$$  
\projection{\atom{p, t}}{}
\selection{\atom{cL} = \atom{pL}}
\unnest{\atom{c.lang} \assign \atom{cL}, \atom{p.lang} \assign \atom{pL}}
\left( 
\getvertices{p}{Post}
\join^*
\left(\getedgesdirected{p}{Post}{c}{Comm}{}{REPLY}\right)
\right)
$$

(3)~Transform NRA to FRA following the approaches presented in~\cite{DBLP:journals/tods/ParedaensG92,DBLP:journals/tcs/Bussche01}. However, a key difference is that due to their schema-free nature, \emph{the schema of the nested relations is not known for property graphs in advance} and has to be inferred based on the query. Therefore, this step includes pushing down nested attributes to the $\getverticesop$ and $\getedgesopdirected$ operators. On the example, this results in
$$  
\projection{\atom{p, t}}{}
\selection{\atom{cL} = \atom{pL}}
\left( 
\getvertices{p}{Post \{lang \assign pL\}}
\join^*
\left(\getedgesdirected{p}{Post}{c}{Comm \{lang \assign cL\}}{}{REPLY}\right)
\right),
$$

where the notation $\{ \atom{lang \assign pL} \}$ represents a property that must be included in the base relation returned by the $\getverticesop$ or $\getedgesopdirected$ operator.

(4)~Create an incremental view for the FRA expression. Incremental view maintenance algorithms for FRA are well studied both from a theoretical perspective~\cite{DBLP:conf/sigmod/BlakeleyLT86,DBLP:conf/sigmod/GuptaMS93,DBLP:conf/sigmod/GriffinL95,BergmannPhD} 
and implementation-wise, with many practical tools~\cite{DBLP:journals/sosym/VarroBHHRU16,drools} and research prototypes~\cite{PerPol2017,DBLP:conf/sigmod/KankanamgeSMCS17,DBLP:journals/pvldb/RenCKLBRZK17}. While they are not expressible in first-order logic, it is possible to evaluate transitive operations incrementally~\cite{DBLP:journals/tods/PangDR05,DBLP:conf/gg/BergmannRSTV12}.

Based on the proposed approach, we state the following: \emph{The \opencypher language with unordered bags (instead of lists) and atomic paths (which can only be inserted or deleted, and lose their ordering when unnested) is incrementally maintainable.}


\myparagraph{Evaluation}

The presented approach allows IVM for property graph queries, while allowing FGN and some degree of ORD (for handling paths). In particular, the proposed fragment  still allows returning paths and \emph{path unwinding}~\cite{DBLP:journals/corr/AnglesABHRV16}, a feature that permits the query to iterate over the nodes of a path variable.
The main tradeoff of the approach is that it does not allow users to use lists in their data model and queries. It is also not possible to specify top-$k$ style queries, \eg get the top 3 messages, based on the number of replies received.

\myparagraph{Summary of contributions}

Up to our best knowledge, our research is the first to investigate challenges of \emph{incremental view maintenance for property graph queries}. We put a particular emphasis on handling nested data structures and ordering; and propose to limit the usage of ordering for (atomic) paths.
Formulating the queries in NRA and flattening it to an FRA expression allows us to infer the \emph{minimal schema} required by each operator, based on the query specification. Our approach does not require a priori knowledge of the data schema, unlike the \emph{schema cleanup} algorithm of~\cite{DBLP:conf/widm/ZhangPR02} (defined in the context of evaluating XQuery expressions on XML documents) and the \emph{schema merging} algorithm of~\cite{DBLP:conf/icde/LiQKGG11} (defined for consolidating multiple schemas into a mediated one).

\myparagraph{Limitations and future work}

Property graph queries present numerous additional challenges that were not presented in this paper. In particular, aggregations, the \lstinline{OPTIONAL MATCH}, \lstinline{WITH}, \lstinline{SKIP} constructs were omitted, and are discussed (for non-incremental queries) in our earlier work~\cite{DBLP:conf/adbis/MartonSV17}. Expressions (arithmetic operations, comparisons, and functions) were also left for future work.

%% file: ms.bbl

\begin{thebibliography}{00}


\ifx \showCODEN    \undefined \def \showCODEN     #1{\unskip}     \fi
\ifx \showDOI      \undefined \def \showDOI       #1{{\tt DOI:}\penalty0{#1}\ }
  \fi
\ifx \showISBNx    \undefined \def \showISBNx     #1{\unskip}     \fi
\ifx \showISBNxiii \undefined \def \showISBNxiii  #1{\unskip}     \fi
\ifx \showISSN     \undefined \def \showISSN      #1{\unskip}     \fi
\ifx \showLCCN     \undefined \def \showLCCN      #1{\unskip}     \fi
\ifx \shownote     \undefined \def \shownote      #1{#1}          \fi
\ifx \showarticletitle \undefined \def \showarticletitle #1{#1}   \fi
\ifx \showURL      \undefined \def \showURL       {\relax}        \fi
\providecommand\bibfield[2]{#2}
\providecommand\bibinfo[2]{#2}
\providecommand\natexlab[1]{#1}
\providecommand\showeprint[2][]{arXiv:#2}

\bibitem[\protect\citeauthoryear{Angles, Arenas, Barcel{\'{o}}, Hogan, Reutter,
  and Vrgoc}{Angles et~al\mbox{.}}{2016}]%
        {DBLP:journals/corr/AnglesABHRV16}
\bibfield{author}{\bibinfo{person}{Renzo Angles}, \bibinfo{person}{Marcelo
  Arenas}, \bibinfo{person}{Pablo Barcel{\'{o}}}, \bibinfo{person}{Aidan
  Hogan}, \bibinfo{person}{Juan~L. Reutter}, {and} \bibinfo{person}{Domagoj
  Vrgoc}.} \bibinfo{year}{2016}\natexlab{}.
\newblock \showarticletitle{Foundations of Modern Graph Query Languages}.
\newblock \bibinfo{journal}{{\em CoRR\/}}  \bibinfo{volume}{abs/1610.06264}
  (\bibinfo{year}{2016}).
\newblock
\showURL{%
\url{http://arxiv.org/abs/1610.06264}}


\bibitem[\protect\citeauthoryear{Bergmann}{Bergmann}{2013}]%
        {BergmannPhD}
\bibfield{author}{\bibinfo{person}{G{\'a}bor Bergmann}.}
  \bibinfo{year}{2013}\natexlab{}.
\newblock {\em \bibinfo{title}{Incremental Model Queries in Model-Driven
  Design}}.
\newblock {Ph.D. dissertation}. \bibinfo{school}{Budapest University of
  Technology and Economics}, \bibinfo{address}{Budapest}.
\newblock


\bibitem[\protect\citeauthoryear{Bergmann, R{\'{a}}th, Szab{\'{o}}, Torrini,
  and Varr{\'{o}}}{Bergmann et~al\mbox{.}}{2012}]%
        {DBLP:conf/gg/BergmannRSTV12}
\bibfield{author}{\bibinfo{person}{G{\'{a}}bor Bergmann},
  \bibinfo{person}{Istv{\'{a}}n R{\'{a}}th}, \bibinfo{person}{Tam{\'{a}}s
  Szab{\'{o}}}, \bibinfo{person}{Paolo Torrini}, {and}
  \bibinfo{person}{D{\'{a}}niel Varr{\'{o}}}.} \bibinfo{year}{2012}\natexlab{}.
\newblock \showarticletitle{Incremental Pattern Matching for the Efficient
  Computation of Transitive Closure}. In \bibinfo{booktitle}{{\em Graph
  Transformations - 6th International Conference, {ICGT} 2012, Bremen, Germany,
  September 24-29, 2012. Proceedings}} {\em (\bibinfo{series}{Lecture Notes in
  Computer Science})}, \bibfield{editor}{\bibinfo{person}{Hartmut Ehrig},
  \bibinfo{person}{Gregor Engels}, \bibinfo{person}{Hans{-}J{\"{o}}rg
  Kreowski}, {and} \bibinfo{person}{Grzegorz Rozenberg}} (Eds.),
  Vol.~\bibinfo{volume}{7562}. \bibinfo{publisher}{Springer},
  \bibinfo{pages}{386--400}.
\newblock
\showDOI{%
\url{https://doi.org/10.1007/978-3-642-33654-6_26}}


\bibitem[\protect\citeauthoryear{Blakeley, Larson, and Tompa}{Blakeley
  et~al\mbox{.}}{1986}]%
        {DBLP:conf/sigmod/BlakeleyLT86}
\bibfield{author}{\bibinfo{person}{Jos{\'{e}}~A. Blakeley},
  \bibinfo{person}{Per{-}{\AA}ke Larson}, {and} \bibinfo{person}{Frank~Wm.
  Tompa}.} \bibinfo{year}{1986}\natexlab{}.
\newblock \showarticletitle{Efficiently Updating Materialized Views}. In
  \bibinfo{booktitle}{{\em {SIGMOD}}}. \bibinfo{pages}{61--71}.
\newblock
\showDOI{%
\url{https://doi.org/10.1145/16894.16861}}


\bibitem[\protect\citeauthoryear{Botoeva et~al\mbox{.}}{Botoeva
  et~al\mbox{.}}{2016a}]%
        {DBLP:journals/corr/BotoevaCCRX16}
\bibfield{author}{\bibinfo{person}{Elena Botoeva} {and}
  \bibinfo{person}{others}.} \bibinfo{year}{2016}\natexlab{a}.
\newblock \showarticletitle{A Formal Presentation of {MongoDB} (Extended
  Version)}.
\newblock \bibinfo{journal}{{\em CoRR\/}}  \bibinfo{volume}{abs/1603.09291}
  (\bibinfo{year}{2016}).
\newblock
\showURL{%
\url{http://arxiv.org/abs/1603.09291}}


\bibitem[\protect\citeauthoryear{Botoeva et~al\mbox{.}}{Botoeva
  et~al\mbox{.}}{2016b}]%
        {DBLP:conf/dlog/BotoevaCCRX16}
\bibfield{author}{\bibinfo{person}{Elena Botoeva} {and}
  \bibinfo{person}{others}.} \bibinfo{year}{2016}\natexlab{b}.
\newblock \showarticletitle{{OBDA} Beyond Relational {DBs}: {A} Study for
  {MongoDB}}. In \bibinfo{booktitle}{{\em Description Logics}}.
\newblock


\bibitem[\protect\citeauthoryear{den Bussche}{den Bussche}{2001}]%
        {DBLP:journals/tcs/Bussche01}
\bibfield{author}{\bibinfo{person}{Jan~Van den Bussche}.}
  \bibinfo{year}{2001}\natexlab{}.
\newblock \showarticletitle{Simulation of the nested relational algebra by the
  flat relational algebra, with an application to the complexity of evaluating
  powerset algebra expressions}.
\newblock \bibinfo{journal}{{\em Theor. Comput. Sci.\/}} \bibinfo{volume}{254},
  \bibinfo{number}{1-2} (\bibinfo{year}{2001}), \bibinfo{pages}{363--377}.
\newblock
\showDOI{%
\url{https://doi.org/10.1016/S0304-3975(99)00301-1}}


\bibitem[\protect\citeauthoryear{Dimitrova, El{-}Sayed, and
  Rundensteiner}{Dimitrova et~al\mbox{.}}{2003a}]%
        {XQueryTechReport}
\bibfield{author}{\bibinfo{person}{Katica Dimitrova}, \bibinfo{person}{Maged
  El{-}Sayed}, {and} \bibinfo{person}{Elke~A. Rundensteiner}.}
  \bibinfo{year}{2003}\natexlab{a}.
\newblock \bibinfo{booktitle}{{\em Order-Sensitive View Maintenance of
  Materialized XQuery Views}}.
\newblock \bibinfo{type}{{T}echnical {R}eport}. \bibinfo{institution}{Computer
  Science Department, Worcester Polytechnic Institute}.
\newblock
\newblock
\shownote{WPI-CS-TR-03-17.}


\bibitem[\protect\citeauthoryear{Dimitrova, El{-}Sayed, and
  Rundensteiner}{Dimitrova et~al\mbox{.}}{2003b}]%
        {DBLP:conf/er/DimitrovaER03}
\bibfield{author}{\bibinfo{person}{Katica Dimitrova}, \bibinfo{person}{Maged
  El{-}Sayed}, {and} \bibinfo{person}{Elke~A. Rundensteiner}.}
  \bibinfo{year}{2003}\natexlab{b}.
\newblock \showarticletitle{Order-Sensitive View Maintenance of Materialized
  XQuery Views}. In \bibinfo{booktitle}{{\em {ER}}}. \bibinfo{pages}{144--157}.
\newblock
\showDOI{%
\url{https://doi.org/10.1007/978-3-540-39648-2_14}}


\bibitem[\protect\citeauthoryear{Griffin and Libkin}{Griffin and
  Libkin}{1995}]%
        {DBLP:conf/sigmod/GriffinL95}
\bibfield{author}{\bibinfo{person}{Timothy Griffin} {and}
  \bibinfo{person}{Leonid Libkin}.} \bibinfo{year}{1995}\natexlab{}.
\newblock \showarticletitle{Incremental Maintenance of Views with Duplicates}.
  In \bibinfo{booktitle}{{\em {SIGMOD}}}. \bibinfo{pages}{328--339}.
\newblock
\showDOI{%
\url{https://doi.org/10.1145/223784.223849}}


\bibitem[\protect\citeauthoryear{Gupta, Mumick, and Subrahmanian}{Gupta
  et~al\mbox{.}}{1993}]%
        {DBLP:conf/sigmod/GuptaMS93}
\bibfield{author}{\bibinfo{person}{Ashish Gupta},
  \bibinfo{person}{Inderpal~Singh Mumick}, {and} \bibinfo{person}{V.~S.
  Subrahmanian}.} \bibinfo{year}{1993}\natexlab{}.
\newblock \showarticletitle{Maintaining Views Incrementally}. In
  \bibinfo{booktitle}{{\em {SIGMOD}}}. \bibinfo{pages}{157--166}.
\newblock
\showDOI{%
\url{https://doi.org/10.1145/170035.170066}}


\bibitem[\protect\citeauthoryear{Hat}{Hat}{2017}]%
        {drools}
\bibfield{author}{\bibinfo{person}{Red Hat}.} \bibinfo{year}{2017}\natexlab{}.
\newblock \bibinfo{title}{Drools}.
\newblock \bibinfo{howpublished}{\url{http://www.drools.org/}}.
  (\bibinfo{year}{2017}).
\newblock


\bibitem[\protect\citeauthoryear{H{\"{o}}lsch and Grossniklaus}{H{\"{o}}lsch
  and Grossniklaus}{2016}]%
        {DBLP:conf/edbt/HolschG16}
\bibfield{author}{\bibinfo{person}{J{\"{u}}rgen H{\"{o}}lsch} {and}
  \bibinfo{person}{Michael Grossniklaus}.} \bibinfo{year}{2016}\natexlab{}.
\newblock \showarticletitle{An Algebra and Equivalences to Transform Graph
  Patterns in {N}eo4j}. In \bibinfo{booktitle}{{\em {GraphQ} at {EDBT/ICDT}}}.
\newblock


\bibitem[\protect\citeauthoryear{Jaeschke and Schek}{Jaeschke and
  Schek}{1982}]%
        {DBLP:conf/pods/JaeschkeS82}
\bibfield{author}{\bibinfo{person}{Gerhard Jaeschke} {and}
  \bibinfo{person}{Hans{-}J{\"{o}}rg Schek}.} \bibinfo{year}{1982}\natexlab{}.
\newblock \showarticletitle{Remarks on the Algebra of Non First Normal Form
  Relations}. In \bibinfo{booktitle}{{\em {PODS}}},
  \bibfield{editor}{\bibinfo{person}{Jeffrey~D. Ullman} {and}
  \bibinfo{person}{Alfred~V. Aho}} (Eds.). \bibinfo{publisher}{{ACM}},
  \bibinfo{pages}{124--138}.
\newblock
\showDOI{%
\url{https://doi.org/10.1145/588111.588133}}


\bibitem[\protect\citeauthoryear{Kankanamge et~al\mbox{.}}{Kankanamge
  et~al\mbox{.}}{2017}]%
        {DBLP:conf/sigmod/KankanamgeSMCS17}
\bibfield{author}{\bibinfo{person}{Chathura Kankanamge} {and}
  \bibinfo{person}{others}.} \bibinfo{year}{2017}\natexlab{}.
\newblock \showarticletitle{Graphflow: An Active Graph Database}. In
  \bibinfo{booktitle}{{\em {SIGMOD}}}. \bibinfo{pages}{1695--1698}.
\newblock
\showDOI{%
\url{https://doi.org/10.1145/3035918.3056445}}


\bibitem[\protect\citeauthoryear{Kuno and Rundensteiner}{Kuno and
  Rundensteiner}{1998}]%
        {DBLP:journals/tkde/KunoR98}
\bibfield{author}{\bibinfo{person}{Harumi~A. Kuno} {and}
  \bibinfo{person}{Elke~A. Rundensteiner}.} \bibinfo{year}{1998}\natexlab{}.
\newblock \showarticletitle{Incremental Maintenance of Materialized
  Object-Oriented Views in MultiView: Strategies and Performance Evaluation}.
\newblock \bibinfo{journal}{{\em {IEEE} Trans. Knowl. Data Eng.\/}}
  \bibinfo{volume}{10}, \bibinfo{number}{5} (\bibinfo{year}{1998}),
  \bibinfo{pages}{768--792}.
\newblock
\showDOI{%
\url{https://doi.org/10.1109/69.729731}}


\bibitem[\protect\citeauthoryear{{LDBC Social Network Benchmark task
  force}}{{LDBC Social Network Benchmark task force}}{2017}]%
        {LDBC_SNB}
\bibfield{author}{\bibinfo{person}{{LDBC Social Network Benchmark task
  force}}.} \bibinfo{year}{2017}\natexlab{}.
\newblock \bibinfo{booktitle}{{\em LDBC Social Network Benchmark (SNB)}}.
\newblock \bibinfo{type}{{T}echnical {R}eport}. \bibinfo{institution}{Linked
  Data Benchmark Council}.
\newblock
\newblock
\shownote{\url{http://https://ldbc.github.io/ldbc_snb_docs/ldbc-snb-specification.pdf}.}


\bibitem[\protect\citeauthoryear{Li, Quix, Kensche, Geisler, and Guo}{Li
  et~al\mbox{.}}{2011}]%
        {DBLP:conf/icde/LiQKGG11}
\bibfield{author}{\bibinfo{person}{Xiang Li}, \bibinfo{person}{Christoph Quix},
  \bibinfo{person}{David Kensche}, \bibinfo{person}{Sandra Geisler}, {and}
  \bibinfo{person}{Lisong Guo}.} \bibinfo{year}{2011}\natexlab{}.
\newblock \showarticletitle{Automatic generation of mediated schemas through
  reasoning over data dependencies}. In \bibinfo{booktitle}{{\em {ICDE}}}.
  \bibinfo{pages}{1280--1283}.
\newblock
\showDOI{%
\url{https://doi.org/10.1109/ICDE.2011.5767913}}


\bibitem[\protect\citeauthoryear{Liu, Vincent, and Mohania}{Liu
  et~al\mbox{.}}{1999}]%
        {DBLP:conf/ideas/LiuVM99}
\bibfield{author}{\bibinfo{person}{Jixue Liu}, \bibinfo{person}{Millist~W.
  Vincent}, {and} \bibinfo{person}{Mukesh~K. Mohania}.}
  \bibinfo{year}{1999}\natexlab{}.
\newblock \showarticletitle{Incremental Maintenance of Nested Relational
  Views}. In \bibinfo{booktitle}{{\em {IDEAS}}}. \bibinfo{pages}{197--205}.
\newblock
\showDOI{%
\url{https://doi.org/10.1109/IDEAS.1999.787268}}


\bibitem[\protect\citeauthoryear{Marton, Sz{\'{a}}rnyas, and
  Varr{\'{o}}}{Marton et~al\mbox{.}}{2017}]%
        {DBLP:conf/adbis/MartonSV17}
\bibfield{author}{\bibinfo{person}{J{\'{o}}zsef Marton},
  \bibinfo{person}{G{\'{a}}bor Sz{\'{a}}rnyas}, {and}
  \bibinfo{person}{D{\'{a}}niel Varr{\'{o}}}.} \bibinfo{year}{2017}\natexlab{}.
\newblock \showarticletitle{Formalising openCypher Graph Queries in Relational
  Algebra}. In \bibinfo{booktitle}{{\em {ADBIS}}}. \bibinfo{pages}{182--196}.
\newblock
\showDOI{%
\url{https://doi.org/10.1007/978-3-319-66917-5_13}}


\bibitem[\protect\citeauthoryear{{Neo Technology}}{{Neo Technology}}{2017}]%
        {openCypher}
\bibfield{author}{\bibinfo{person}{{Neo Technology}}.}
  \bibinfo{year}{2017}\natexlab{}.
\newblock \bibinfo{title}{open{C}ypher Project}.
\newblock \bibinfo{howpublished}{\url{http://www.opencypher.org/}}.
  (\bibinfo{year}{2017}).
\newblock


\bibitem[\protect\citeauthoryear{openCypher}{openCypher}{2017}]%
        {CAPS}
\bibfield{author}{\bibinfo{person}{openCypher}.}
  \bibinfo{year}{2017}\natexlab{}.
\newblock \bibinfo{title}{{CAPS}: {C}ypher for {A}pache {S}park}.
\newblock
  \bibinfo{howpublished}{\url{https://github.com/opencypher/cypher-for-apache-spark}}.
    (\bibinfo{year}{2017}).
\newblock


\bibitem[\protect\citeauthoryear{Pang, Dong, and Ramamohanarao}{Pang
  et~al\mbox{.}}{2005}]%
        {DBLP:journals/tods/PangDR05}
\bibfield{author}{\bibinfo{person}{Chaoyi Pang}, \bibinfo{person}{Guozhu Dong},
  {and} \bibinfo{person}{Kotagiri Ramamohanarao}.}
  \bibinfo{year}{2005}\natexlab{}.
\newblock \showarticletitle{Incremental maintenance of shortest distance and
  transitive closure in first-order logic and {SQL}}.
\newblock \bibinfo{journal}{{\em {ACM} Trans. Database Syst.\/}}
  \bibinfo{volume}{30}, \bibinfo{number}{3} (\bibinfo{year}{2005}),
  \bibinfo{pages}{698--721}.
\newblock
\showDOI{%
\url{https://doi.org/10.1145/1093382.1093384}}


\bibitem[\protect\citeauthoryear{Paradies et~al\mbox{.}}{Paradies
  et~al\mbox{.}}{2017}]%
        {DBLP:conf/dbpl/ParadiesKBFKG17}
\bibfield{author}{\bibinfo{person}{Marcus Paradies} {and}
  \bibinfo{person}{others}.} \bibinfo{year}{2017}\natexlab{}.
\newblock \showarticletitle{GraphScript: implementing complex graph algorithms
  in {SAP} {HANA}}. In \bibinfo{booktitle}{{\em DBPL}}.
  \bibinfo{pages}{13:1--13:4}.
\newblock
\showDOI{%
\url{https://doi.org/10.1145/3122831.3122841}}


\bibitem[\protect\citeauthoryear{Paredaens and Gucht}{Paredaens and
  Gucht}{1992}]%
        {DBLP:journals/tods/ParedaensG92}
\bibfield{author}{\bibinfo{person}{Jan Paredaens} {and}
  \bibinfo{person}{Dirk~Van Gucht}.} \bibinfo{year}{1992}\natexlab{}.
\newblock \showarticletitle{Converting Nested Algebra Expressions into Flat
  Algebra Expressions}.
\newblock \bibinfo{journal}{{\em {ACM} Trans. Database Syst.\/}}
  \bibinfo{volume}{17}, \bibinfo{number}{1} (\bibinfo{year}{1992}),
  \bibinfo{pages}{65--93}.
\newblock
\showDOI{%
\url{https://doi.org/10.1145/128765.128768}}


\bibitem[\protect\citeauthoryear{Ren et~al\mbox{.}}{Ren et~al\mbox{.}}{2017}]%
        {DBLP:journals/pvldb/RenCKLBRZK17}
\bibfield{author}{\bibinfo{person}{Xiangnan Ren} {and}
  \bibinfo{person}{others}.} \bibinfo{year}{2017}\natexlab{}.
\newblock \showarticletitle{Strider: An Adaptive, Inference-enabled Distributed
  {RDF} Stream Processing Engine}.
\newblock \bibinfo{journal}{{\em {PVLDB}\/}} \bibinfo{volume}{10},
  \bibinfo{number}{12} (\bibinfo{year}{2017}), \bibinfo{pages}{1905--1908}.
\newblock
\showURL{%
\url{http://www.vldb.org/pvldb/vol10/p1905-ren.pdf}}


\bibitem[\protect\citeauthoryear{Robinson, Webber, and Eifrém}{Robinson
  et~al\mbox{.}}{2015}]%
        {GraphDatabases}
\bibfield{author}{\bibinfo{person}{Ian Robinson}, \bibinfo{person}{Jim Webber},
  {and} \bibinfo{person}{Emil Eifrém}.} \bibinfo{year}{2015}\natexlab{}.
\newblock \bibinfo{booktitle}{{\em Graph Databases\/} (\bibinfo{edition}{2nd}
  ed.)}.
\newblock \bibinfo{publisher}{O'Reilly Media}.
\newblock
\showISBNx{9781491930892}


\bibitem[\protect\citeauthoryear{Rodriguez}{Rodriguez}{2015}]%
        {DBLP:conf/dbpl/Rodriguez15}
\bibfield{author}{\bibinfo{person}{Marko~A. Rodriguez}.}
  \bibinfo{year}{2015}\natexlab{}.
\newblock \showarticletitle{The Gremlin graph traversal machine and language
  (invited talk)}. In \bibinfo{booktitle}{{\em DBPL}}. \bibinfo{pages}{1--10}.
\newblock
\showDOI{%
\url{https://doi.org/10.1145/2815072.2815073}}


\bibitem[\protect\citeauthoryear{Sahu, Mhedhbi, Salihoglu, Lin, and
  {\"{O}}zsu}{Sahu et~al\mbox{.}}{2017}]%
        {DBLP:journals/corr/abs-1709-03188}
\bibfield{author}{\bibinfo{person}{Siddhartha Sahu}, \bibinfo{person}{Amine
  Mhedhbi}, \bibinfo{person}{Semih Salihoglu}, \bibinfo{person}{Jimmy Lin},
  {and} \bibinfo{person}{M.~Tamer {\"{O}}zsu}.}
  \bibinfo{year}{2017}\natexlab{}.
\newblock \showarticletitle{The Ubiquity of Large Graphs and Surprising
  Challenges of Graph Processing: {A} User Survey}.
\newblock \bibinfo{journal}{{\em CoRR\/}}  \bibinfo{volume}{abs/1709.03188}
  (\bibinfo{year}{2017}).
\newblock
\showURL{%
\url{http://arxiv.org/abs/1709.03188}}


\bibitem[\protect\citeauthoryear{Sz{\'a}rnyas, Izs{\'o}, R{\'a}th, and
  Varr{\'o}}{Sz{\'a}rnyas et~al\mbox{.}}{2017a}]%
        {TrainBenchmarkSOSYM}
\bibfield{author}{\bibinfo{person}{G{\'a}bor Sz{\'a}rnyas},
  \bibinfo{person}{Benedek Izs{\'o}}, \bibinfo{person}{Istv{\'a}n R{\'a}th},
  {and} \bibinfo{person}{D{\'a}niel Varr{\'o}}.}
  \bibinfo{year}{2017}\natexlab{a}.
\newblock \showarticletitle{The {T}rain {B}enchmark: Cross-Technology
  Performance Evaluation of Continuous Model Validation}.
\newblock \bibinfo{journal}{{\em Softw. Syst. Model.\/}}
  (\bibinfo{year}{2017}).
\newblock


\bibitem[\protect\citeauthoryear{Sz{\'a}rnyas, Maginecz, and
  Varr{\'o}}{Sz{\'a}rnyas et~al\mbox{.}}{2017b}]%
        {PerPol2017}
\bibfield{author}{\bibinfo{person}{G{\'a}bor Sz{\'a}rnyas},
  \bibinfo{person}{J{\'a}nos Maginecz}, {and} \bibinfo{person}{D{\'a}niel
  Varr{\'o}}.} \bibinfo{year}{2017}\natexlab{b}.
\newblock \showarticletitle{Evaluation of Optimization Strategies Incremental
  Graph Query Evaluation}.
\newblock \bibinfo{journal}{{\em Periodica Polytechnica, Electrical Engineering
  and Computer Science\/}} (\bibinfo{year}{2017}).
\newblock


\bibitem[\protect\citeauthoryear{Ujhelyi et~al\mbox{.}}{Ujhelyi
  et~al\mbox{.}}{2015}]%
        {DBLP:journals/infsof/UjhelyiSHCVVF15}
\bibfield{author}{\bibinfo{person}{Zolt{\'{a}}n Ujhelyi} {and}
  \bibinfo{person}{others}.} \bibinfo{year}{2015}\natexlab{}.
\newblock \showarticletitle{Performance comparison of query-based techniques
  for anti-pattern detection}.
\newblock \bibinfo{journal}{{\em Information {\&} Software Technology\/}}
  \bibinfo{volume}{65} (\bibinfo{year}{2015}), \bibinfo{pages}{147--165}.
\newblock
\showDOI{%
\url{https://doi.org/10.1016/j.infsof.2015.01.003}}


\bibitem[\protect\citeauthoryear{Varr{\'{o}}, Bergmann, Heged{\"{u}}s,
  Horv{\'{a}}th, R{\'{a}}th, and Ujhelyi}{Varr{\'{o}} et~al\mbox{.}}{2016}]%
        {DBLP:journals/sosym/VarroBHHRU16}
\bibfield{author}{\bibinfo{person}{D{\'{a}}niel Varr{\'{o}}},
  \bibinfo{person}{G{\'{a}}bor Bergmann}, \bibinfo{person}{{\'{A}}bel
  Heged{\"{u}}s}, \bibinfo{person}{{\'{A}}kos Horv{\'{a}}th},
  \bibinfo{person}{Istv{\'{a}}n R{\'{a}}th}, {and}
  \bibinfo{person}{Zolt{\'{a}}n Ujhelyi}.} \bibinfo{year}{2016}\natexlab{}.
\newblock \showarticletitle{Road to a reactive and incremental model
  transformation platform: three generations of the {VIATRA} framework}.
\newblock \bibinfo{journal}{{\em Softw. Syst. Model.\/}} \bibinfo{volume}{15},
  \bibinfo{number}{3} (\bibinfo{year}{2016}), \bibinfo{pages}{609--629}.
\newblock
\showDOI{%
\url{https://doi.org/10.1007/s10270-016-0530-4}}


\bibitem[\protect\citeauthoryear{Zhang, Pielech, and Rundensteiner}{Zhang
  et~al\mbox{.}}{2002}]%
        {DBLP:conf/widm/ZhangPR02}
\bibfield{author}{\bibinfo{person}{Xin Zhang}, \bibinfo{person}{Bradford
  Pielech}, {and} \bibinfo{person}{Elke~A. Rundensteiner}.}
  \bibinfo{year}{2002}\natexlab{}.
\newblock \showarticletitle{Honey, {I} shrunk the {XQuery}!: an {XML} algebra
  optimization approach}. In \bibinfo{booktitle}{{\em {WIDM} at {CIKM}}}.
  \bibinfo{pages}{15--22}.
\newblock
\showDOI{%
\url{https://doi.org/10.1145/584931.584936}}


\end{thebibliography}
